\documentclass[amsmath,amssymb,aps,reprint,nofootinbib,hidelinks,superscriptaddress]{revtex4-2}

\usepackage{hyperref} 
\usepackage{graphicx}
\usepackage{dcolumn}
\usepackage{bm}
\usepackage{graphicx}
\usepackage{epstopdf}
\usepackage{mathrsfs}
\usepackage{amssymb}
\usepackage{slashed}
\usepackage{verbatim}
\usepackage{color}
\usepackage{multirow}
\usepackage{amsmath}
\usepackage{mathrsfs}
\usepackage{epsfig}
\usepackage{ulem}
\usepackage{graphics}
\usepackage{indentfirst}
\usepackage{slashed}

\definecolor{cardinal}{rgb}{0.6,0,0}
\definecolor{darkgreen}{rgb}{0,0.5,0}
\definecolor{golden}{rgb}{0.92, 0.7, 0}
\definecolor{midnight}{rgb}{0, 0, 0.5}
\definecolor{darkblue}{rgb}{0.2, 0, 0.8}

\newcommand{\Lgr}{{\mathscr{L}}}
\newcommand{\Veff}{{\mathscr{V}_{\rm eff}}}
\newcommand{\covD}{{\mathscr{D}}}

\begin{document}

\title{
Vacuum stability, fixed points, and phases of QED$_3$ at large $N_f$}

\author{Lorenzo Di Pietro} 
\affiliation{Dipartimento di Fisica, Universit\`{a} di Trieste, Strada Costiera 11, I-34151 Trieste, Italy,\\ \& INFN, Sezione di Trieste, Via Valerio 2, I-34127 Trieste, Italy}
\author{Edoardo Lauria}
\affiliation{CPHT, CNRS, Institut Polytechnique de Paris, France}
\author{Pierluigi Niro} 
\affiliation{Mani L. Bhaumik Institute for Theoretical Physics, Department of Physics and Astronomy, University of California, Los Angeles, CA 90095, USA}

\begin{abstract}
\noindent We consider three-dimensional Quantum Electrodynamics in the presence of a Chern-Simons term at level $k$ and $N_f$ flavors, in the limit of large $N_f$ and $k$ with $k/N_f$ fixed. We consider either bosonic or fermionic matter fields, with and without quartic terms at criticality: the resulting theories are critical and tricritical bosonic QED$_3$, Gross-Neveu and fermionic QED$_3$.
For all such theories we compute the effective potentials and the $\beta$ functions of classically marginal couplings, at the leading order in the large $N_f$ limit and to all orders in $k/N_f$ and in the couplings. We determine the RG fixed points and discuss the quantum stability of the corresponding vacua.
While critical bosonic and fermionic QED$_3$ are always stable CFTs, we find that tricritical bosonic and Gross-Neveu QED$_3$ exist as stable CFTs only for specific values of $k/N_f$. Finally, we discuss the phase diagrams of these theories as a function of their relevant deformations.
\end{abstract}

\maketitle

\section{Introduction}
\label{sec:introduction}
Three-dimensional Quantum Electrodynamics (QEDs) with either bosonic or fermionic degrees of freedom are among the simplest and yet very rich examples of gauge theories. At high energies, these theories are defined in terms of a three-dimensional $U(1)$ Maxwell field $a$ with gauge coupling $e^2$ (of mass dimension one) and a Chern-Simons term at level $k$, coupled to $N_f$ flavors of charged bosons or fermions. The Lagrangian is (we will always work in Euclidean signature)
\begin{equation}\label{LagrGen}
\Lgr =  \frac{1}{2e^2} da \wedge \star da + \frac{i k}{4\pi} a \wedge da +\Lgr_{\text{matter}}\,.
\end{equation}
In absence of charged matter, if $k=0$ the theory is dual to a compact scalar and it confines in the sense of \cite{Polyakov:1975rs,Polyakov:1976fu}, whereas if $k\neq 0$ it flows to a pure $U(1)_k$ Chern-Simons theory.
With charged matter, the low-energy behavior of QED theories for small $N_f$ has not been rigorously established, and this question remains an open problem which has been studied with a variety of approaches (see e.g.~refs.~\cite{Appelquist:1988sr, Nash:1989xx, Hong:1992ww, Kondo:1994cz, Maris:1996zg, Fischer:2004nq, Kotikov:2016wrb, Gusynin:2016som, Kubota:2001kk, Kaveh:2004qa, Herbut:2016ide, Braun:2014wja, Giombi:2015haa, DiPietro:2015taa, DiPietro:2017kcd, Zerf:2018csr, Giombi:2016fct, Benvenuti:2018cwd, Benvenuti:2019ujm, Halperin:1973jh, PhysRevB.41.4083, Ihrig:2019kfv, Metayer:2022wre, Chester:2021drl, Chester:2017vdh, Chester:2022wur}), including lattice simulations (see e.g.~refs.~\cite{Hands:2002dv, Hands:2002qt, Hands:2004bh, Strouthos:2007stc, Karthik:2015sgq, Karthik:2016ppr, Bonati:2020jlm}) and conformal bootstrap (see e.g.~\cite{Chester:2016wrc,Chester:2017vdh,Li:2018lyb, Li:2020bnb, He:2021xvg,He:2021sto,Albayrak:2021xtd,Li:2021emd, Manenti:2021elk, Poland:2018epd} and references therein).
When $N_f$ is large enough, these theories are generically expected to flow to interacting Conformal Field Theories (CFTs) at low energies. This expectation is corroborated by the existence of fixed points of the Renormalization Group (RG) flow, which can be found in the limit of large $N_f$.
However, as we will discuss in this work, some of these fixed points happen to lie in a region of instability of the theory.\footnote{In this Letter by ``stability'' we refer to the condition that the potential has a stable minimum, not to the RG stability of the fixed point under deformations by some coupling.}

In order to clarify this issue, we will compute the effective potentials $\Veff$ for a few instances of such QEDs (to be defined below): two bosonic, called `tricritical' and `critical' QED$_3$, and two fermionic, called `Gross-Neveu' and `fermionic' QED$_3$. While critical and fermionic QED$_3$ do not admit any marginal deformations, tricritical and Gross-Neveu QED$_3$ admit classically marginal couplings. Requiring that the quantum vacuum is stable constrains such couplings. 
We will perform these computations by working at the leading order in the $1/N_f$ expansion, with $\Lambda = e^2 N_f$ and $\kappa = k/N_f$ held fixed.
The low-energy limit is reached by taking $\Lambda\rightarrow\infty$ while retaining the dimensionless parameter $\kappa$ which does not run.
The gauge interactions do not enter $\Veff$ at the leading order in the large-$N_f$ limit, nevertheless the parameter $\kappa$ will play an important role in the following analysis.\footnote{The kinetic term does generically affect the vacuum structure of gauge theories with {\it finite} $N_f$, e.g.~it distinguishes between type I and type II superconductivity in the 3d Abelian Higgs model, and it affects the existence of a conformal critical point in large-$N$ $SU(N)_k$ QCD$_3$, as a function of $k/N$~\cite{Armoni:2019lgb}. However, for large-$N_f$ QEDs, the whole $\Lambda$-dependence is encoded in the exact photon propagator and the limit $\Lambda \to \infty$ can be taken reliably.}

Next, we will compute the $\beta$ functions of classically marginal couplings of tricritical and Gross-Neveu QEDs, at leading order in the $1/N_f$ expansion and to all orders in $\kappa$ and in the couplings. 
Details of the RG computations are contained in \cite{DiPietro:2023gzi}. The zeroes of the $\beta$ functions determine families of RG fixed points parametrized by $\kappa$. We will finally impose on those fixed points the constraint of vacuum stability from $\Veff$. The result of this analysis is shown in figures~\ref{hfixedpoints} and~\ref{yfixedpoints}, which summarize the main results of this Letter.

A general takeaway of our analysis is the importance of a joint study of the zeroes of the $\beta$ function {\it and} of the stability of the potential when looking for perturbative fixed points. Another example where this joint analysis is important are Banks-Zaks fixed points in 4d gauge theories, see e.g.~\cite{Benini:2019dfy}.

\section{Bosonic theories}
\label{sec:bosons}

\subsection*{Tricritical QED}
The first theory we consider is tricritical QED$_3$, the theory of $N_f$ massless complex scalars $\phi^m$ ($m=1,\dots,N_f$) coupled to $U(1)_k$ and with the quartic coupling tuned to zero. The matter Lagrangian is
\begin{equation}\label{LmatterTricQED}
\Lgr_{\rm matter}= (\covD_\mu \phi^m)^\dagger(\covD_\mu \phi^m) + \frac{h}{N_f^2}\,(\phi^{\dagger m} \phi^m )^3 \,,
\end{equation}
where $\covD_\mu = \partial_\mu + i a_\mu$ denotes the covariant derivative and $h$ is held fixed in the large-$N_f$ limit. The continuous part of the global symmetry is $SU(N_f) \times U(1)_{\rm m}$, where the first factor is a flavor symmetry and the second one is the magnetic symmetry of the gauge field. For $\kappa=0$ the theory further enjoys parity symmetry. In the $\kappa\rightarrow\infty$ limit the gauge field decouples and the theory describes $2N_f$ real scalars with a sextic interaction, restricted to the $U(1)$-invariant sector. 

Our first task is to study the vacuum stability of tricritical QED$_3$. To this end, we shall compute the effective potential of the theory, at the leading order in the $1/N_f$ expansion and exactly in $\kappa$ and $h$. Following the standard strategy \cite{Moshe:2003xn}, we rewrite the sextic interaction in \eqref{LmatterTricQED} in terms of two auxiliary fields $\sigma$ and $\rho$
\begin{equation}
\sigma \left( \frac{\phi^{\dagger m} \phi^m}{\sqrt{N_f}} - \rho\right) + \frac{h}{\sqrt{N_f}}\,\rho^3 \,.
\end{equation}
Here $\sigma$ is a Lagrange multiplier, while $\rho$ is identified with the composite operator $\phi^{\dagger m} \phi^m/\sqrt{N_f}$. Then, we let $\phi^m = \sqrt{N_f}v^m +\delta\phi^m$, $\rho=\sqrt{N_f}\eta +\delta\rho$, and $\sigma=\sqrt{N_f}\Sigma +\delta\sigma$, being $v^m$, $\eta$, and $\Sigma$ vacuum expectation values (v.e.v.)~that scale as $\mathcal{O}(N_f^0)$. Finally we path-integrate over the fluctuations $\delta\phi^m$, $\delta\rho$, and $\delta\sigma$ to get (in dimensional regularization)
\begin{equation}\label{VeffTric}
	\Veff (v^2,\Sigma,\eta) = N_f \left( \Sigma v^2 - \Sigma \eta + h \eta^3 - \frac{1}{6\pi} \Sigma^{\frac{3}{2}} \right) \,,
\end{equation}
where $v^2 = v^m v^m \geq 0$ (without loss of generality we take $v^m=v\delta^{m1}$), and we require $\Sigma \geq 0$. The derivatives of the potential w.r.t. $v$, $\Sigma$, and $\eta$ are
\begin{equation}
2 \Sigma v = 0 \,, \quad v^2 - \eta - \frac{\sqrt{\Sigma}}{4\pi} = 0 \,, \quad -\Sigma+3h\eta^2=0 \,.
\end{equation}
We refer to these as ``gap equations'' for $v$, $\Sigma$, and $\eta$, respectively. As expected from the absence of scales in the problem, the only stationary point is at $(\Sigma,v,\eta)=P^*\equiv(0,0,0)$. Note that if we restrict to the gap equations for $v$ and $\Sigma$, we find two classes of solutions:
\begin{align}
\begin{split}
  v\neq 0\,,&\quad\Sigma= 0\,,\quad \eta=v^2 > 0~,\hspace{1.3cm}\text{(Higgsed),}~\\
  v=0\,,&\quad\Sigma\geq 0\,, \quad \eta=-\frac{\sqrt\Sigma}{4\pi} \leq 0~,\hspace{0.35cm}\text{(unHiggsed).} \label{directions}
\end{split}
\end{align}
 
Let us now discuss quantum stability. Unlike $\rho$, the auxiliary field $\sigma$ does not correspond on-shell to a physical operator of the theory. As a result we do not study the stability as a function of the expectation value $\Sigma$, but rather we ``integrate it out'' by plugging its gap equation back in the effective potential. Note that the existence of a solution $\Sigma \geq 0$ imposes the constraint $\eta\leq v^2$ on the remaining variables. In this way we express $\Veff$ as
\begin{equation}
\Veff(v^2,\eta) = N_f \left( \frac{16\pi^2}{3}\left(v^2-\eta\right)^3+h\eta^3\right) \,,\, \eta \leq v^2 \,.
\end{equation}
This potential has a global minimum in $P^*$ if\footnote{This result is consistent with ref.~\cite{Aharony:2018pjn}, where the $\kappa\rightarrow\infty$ limit of the stability bound can be recovered as a particular case of their eq.~(1.10), namely in the limit $\lambda_B \rightarrow 0$. We find perfect agreement using the dictionary $4\pi^2 \lambda_B^2 x_6 = h$. In addition, a discussion on the stability of the multicritical point in the large $N$ bosonic vector model can also be found in section 10.2 of ref.~\cite{Moshe:2003xn}. Our stability bound agrees with eq.~(10.22) therein.} 
\begin{equation}
	0 < h < \frac{16\pi^2}{3} \,.
	\label{scalar_bound}
\end{equation} 
Note that, while classically $\eta \geq 0$ and so $h \geq 0$, quantum mechanically $\eta$ can be negative -- the only constraint being $\eta\leq v^2$ -- and the stability region is restricted to eq.~\eqref{scalar_bound}. In other words, bosonic self-interactions are repulsive and tend to destabilize the vacuum. In the appendix, we discuss the effective potential and the phases of the theory in presence of a massive and a quartic deformation, showing how the same bound on $h$ holds.

As a consistency check, note that the determinant of the Hessian $H$ of the effective potential in eq.~\eqref{VeffTric}
\begin{equation}
\det H= - \left( \frac{\sqrt{\Sigma}}{4\pi} + 4v^2 \right) 6 h \eta - 2 \Sigma \,,
\label{hessiandet}
\end{equation}
when restricted to the Higgsed and unHiggsed directions of eq.~\eqref{directions} reads, respectively
\begin{equation}
\begin{split}
\det H_{\rm H} = -24 h \eta^2 \,, \quad \det H_{\rm uH} = (6h-32\pi^2)\eta^2 \,.
\end{split}
\end{equation}
These quantities are both negative everywhere (except $P^*$) precisely when the bound in eq.~\eqref{scalar_bound} is satisfied. This has to be the case since a stable quantum vacuum $P^*$ should be at the same time a minimum of $\Veff$ in two directions and a maximum in the third direction. Indeed, while in the path integrals for $\delta\phi^m$ and $\delta\rho$ the integration contours run along the real axis, in the one for $\delta\sigma$ it runs along the imaginary axis. Correspondingly, the Hessian of $\Veff$ must have two positive and one negative eigenvalues in the neighborhood of $P^*$ and its determinant should be negative. 

Our next task is to verify whether the condition shown in eq.~\eqref{scalar_bound} is satisfied at the RG fixed points of tricritical QED$_3$. To this end we shall compute the $\beta$ function for $h$. At the leading order in the $1/N_f$ expansion and to all orders in $\kappa$ and $h$, the result is \cite{DiPietro:2020fya,DiPietro:2023gzi} 
\begin{align}
\begin{split}
	& \beta_h(h,\kappa)  = \frac{1}{\pi^2 N_f} \left(
	-\frac{9}{256}h^3
	+\frac{9}{4}h^2 \right. \\
	& \hspace{-0.1cm}+\frac{128\pi^2(\pi^2-128\kappa^2)}{(\pi^2+64\kappa^2)^2}h 
	\left. -\frac{16384\pi^4(\pi^2-192\kappa^2)}{3(\pi^2+64\kappa^2)^3}\right) \,.
	\label{betah}
\end{split}	
\end{align}
This $\beta$ function is a cubic polynomial in $h$ and gauge interactions only affect the linear and constant terms. Furthermore, it is an even function of $\kappa$, so we can assume $\kappa>0$ while solving for its zeros.

Let us first inspect the zeros of $\beta_h$ for $\kappa\rightarrow\infty$. In this limit eq.~\eqref{betah} reproduces the known results for the $\beta$ function of the sextic coupling in the free ungauged $O(N)$ model from ref.~\cite{PhysRevLett.48.574}, i.e.\footnote{In particular, our result \eqref{pisarski} matches eq.~(2a) of~\cite{PhysRevLett.48.574}, using the dictionary $N=2N_f$ and $8\pi^2 \lambda=3hN_f^2$.}
\begin{equation}
\beta_h(h,\infty) = \frac{1}{\pi^2 N_f} \left(
-\frac{9}{256}h^3
+\frac{9}{4}h^2\right) \,.
\label{pisarski}
\end{equation}
Hence, we find a double zero at $h^*=0$ and a single zero at $h^*=64$. The sextic operator $(\phi^\dagger \phi)^3$ has dimension exactly $3$ in the free CFT at $h^*=0$, and is marginally irrelevant for $h>0$ and marginally relevant for $h<0$. On the other hand we have $\partial_h\beta_h(h^*=64,\infty)<0$ making the operator relevant there. Note that $h^*=64$ is outside the window \eqref{scalar_bound} of vacuum stability, while $h^*=0$ is at the boundary of the window, and the corresponding theory has a stable vacuum since it is a free CFT.

As a next step, we shall inspect the zeros of $\beta_h$ for $\kappa =0$. In this limit eq.~\eqref{betah} becomes
\begin{align}
\begin{split}
	\label{betahkappa0}
\beta_h(h,0) & = \frac{1}{\pi^2 N_f} \left(
-\frac{9}{256}h^3 \right.\\
&\hspace{2.4cm} \left. +\frac{9}{4}h^2 + 128 \, h - \frac{16384}{3}\right) \,,
\end{split}
\end{align}
and therefore we find one fixed point with irrelevant $(\phi^\dagger \phi)^3$ at $h^*={32(\sqrt{17}-1)}/{3}$ and two fixed points with relevant $(\phi^\dagger \phi)^3$ at $h^*={256}/{3}$ and at $h^*={-32(\sqrt{17}+1)}/{3}$. Note that only the fixed point at $h^*={32(\sqrt{17}-1)}/{3}$ satisfies the vacuum stability bound \eqref{scalar_bound}. Having two singlet relevant deformations, $\phi^\dagger \phi$ and $(\phi^\dagger \phi)^2$, this CFT can be identified with tricritical QED$_3$ at large $N_f$ and $k=0$.

\begin{figure}
	\centering
	\hspace{-0cm}\includegraphics[width=0.5\textwidth]{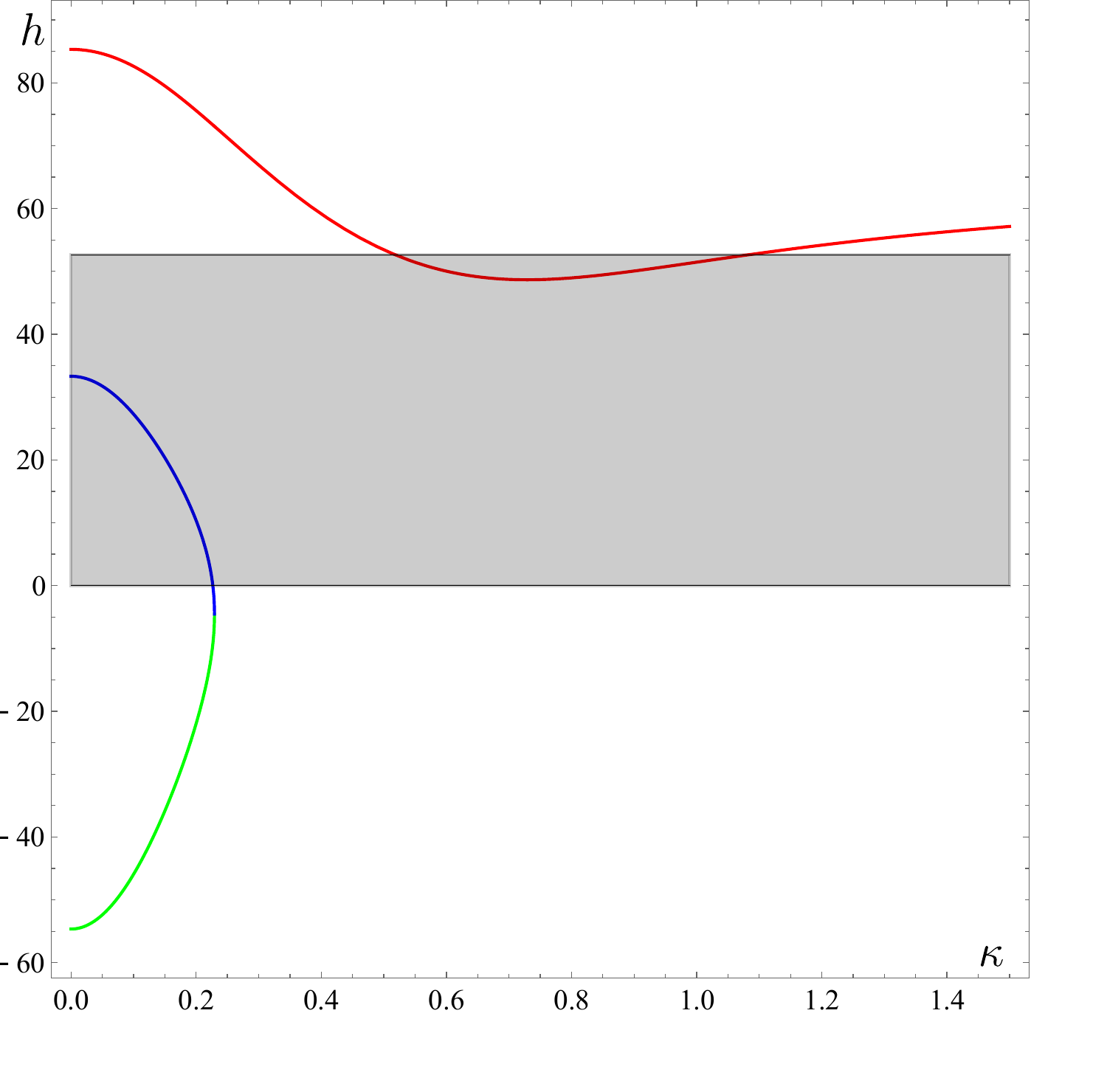}
	\caption{Families of fixed points of $\beta_h$ as a function of $\kappa$ for tricritical QED$_3$. The fixed points corresponding to stable vacua must lie inside the gray region given by $0 < h < \frac{16\pi^2}{3}$.}
	\label{hfixedpoints}
\end{figure}

Besides these special values of $\kappa$, we can solve numerically for the zeros of $\beta_h$ and plot them as functions of $\kappa$, as shown in figure \ref{hfixedpoints}. We see that there are three families of solutions. A family of zeros with relevant $(\phi^\dagger\phi)^3$ (depicted in red in the figure) that exists for any value of $\kappa\geq 0$, and two families of zeros with $(\phi^\dagger\phi)^3$ irrelevant/relevant (depicted in blue/green, respectively) that exist for $\kappa \leq \kappa_0 \simeq 0.229$, above which they annihilate and move to the complex plane, until they reappear at $(\kappa,h)=(\infty,0)$. In the figure, the region highlighted in gray corresponds to the values of $h$ satisfying the vacuum stability bound in eq.~\eqref{scalar_bound}. The blue curve is inside the stability region for $\kappa$ between $\kappa=0$ (where $h^*=32(\sqrt{17}-1)/3$) and $\kappa={\pi}/{8\sqrt{3}} \simeq 0.227$ (where $h^*=0$). The red curve is inside the stability region in the interval $0.518 \lesssim \kappa \lesssim 1.082$. Outside these values of $\kappa$ there is no CFT with a stable vacuum. Note that only the family of fixed points corresponding to the blue curve properly deserves the name of ``tricritical QED'' since the sextic operator is irrelevant there, so the second order transition is reached by tuning the mass and the quartic coupling, while also the sextic coupling requires tuning to reach the family of fixed points on the red curve.

From the results of refs.~\cite{DiPietro:2020fya,DiPietro:2023gzi}  we can also extract the anomalous dimension of the lowest-lying singlet $\phi^{\dagger m} \phi^m$, at the leading order in $1/N_f$ and to all orders in $\kappa$
\begin{equation}
	\gamma_{\phi^\dagger \phi} = \frac{128}{3N_f}\frac{\pi^2-128\kappa^2}{(\pi^2+64\kappa^2)^2} \,,
	\label{anomalous_dimension_free}
\end{equation}
which does not depend on $h$ at this order of the large $N_f$ expansion. As a consistency check, note that for $\kappa=0$ we recover the results of ref.~\cite{Benvenuti:2019ujm} for tricritical QED$_3$, while for $\kappa\rightarrow\infty$ the anomalous dimension vanishes, as we expect since the theory is simply the one of $2N_f$ free real scalars.

\subsection*{Critical QED}

The second theory we consider is critical QED$_3$, where the mass of the $N_f$ complex scalars is still tuned to zero, but the quartic coupling $\lambda$ is at its non-trivial critical point, i.e.~$\lambda=\infty$. The matter Lagrangian is
\begin{equation}
\Lgr_{\rm matter}= (\covD_\mu \phi^m)^\dagger(\covD_\mu \phi^m) + \frac{\sigma}{\sqrt{N_f}}\,\phi^{\dagger m}\phi^m \,,
\label{critqedlagrangian}
\end{equation}
where $\covD_\mu = \partial_\mu + i a_\mu$ is the covariant derivative and $\sigma$ is the Hubbard-Stratonovich (HS) field with scaling dimension $2+\mathcal{O}(1/N_f)$. We did not include a sextic term in the Lagrangian, as from the equation of motion of $\sigma$ we get that $\phi^{\dagger m} \phi^m=0$.
As in the case of tricritical QED$_3$, the continuous global symmetry is $SU(N_f) \times U(1)_{\rm m}$ and the theory further enjoys parity symmetry when $\kappa=0$. In the limit $\kappa\rightarrow \infty$ the gauge field decouples and theory becomes the critical $O(2N_f)$ vector model restricted to the $U(1)$-singlet sector. As it turns out, at the leading order in the $1/N_f$ expansion, critical QED$_3$ exists as a stable CFT for any value of $\kappa$. One can verify this by computing the effective potential $\Veff$ of the theory at large $N_f$. Following a similar strategy as in the case of tricritical QED$_3$, we write $\phi^m=\sqrt{N_f}v^m + \delta\phi^m$ and $\sigma = \sqrt{N_f} \Sigma +\delta\sigma$, and we path-integrate over the fluctuations $\delta\phi^m$ and $\delta\sigma$ to get
\begin{equation}
	\Veff(v^2,\Sigma) = N_f \left( \Sigma v^2 - \frac{1}{6\pi}\Sigma^{\frac{3}{2}} \right) \,,
	\label{effective potential O(N)}
\end{equation}
with again the condition $\Sigma \geq 0$. The gap equations for $v$ and $\Sigma$ read, respectively
\begin{equation}
2 \Sigma v = 0 \,, \quad v^2 - \frac{\sqrt{\Sigma}}{4\pi} = 0 \,,
\end{equation}
which imply that the vacuum is $(v,\Sigma)=P^*\equiv(0,0)$. In order for $P^*$ to be a stable vacuum, we require that, in the neighborhood of $P^*$, one eigenvalue of the Hessian is positive and one is negative. This is equivalent to
\begin{equation}
	\det H
	= - \frac{\sqrt{\Sigma}}{4\pi} - 4v^2 < 0\,,
\end{equation}
which is satisfied everywhere (except at most $P^*$). Since the eigenvalues never change sign, this condition turns out to be also sufficient and we conclude that the vacuum of critical QED$_3$ is always stable for any value of $\kappa$.

From the results of refs.~\cite{DiPietro:2020fya,DiPietro:2023gzi}  we can also extract the anomalous dimension of the lowest-lying singlet $\sigma$, at the leading order in $1/N_f$ and to all orders in $\kappa$
\begin{equation}
	\gamma_\sigma = 
	- \frac{16}{3\pi^2 N_f} \frac{9\pi^4-896\pi^2\kappa^2+4096\kappa^4}{ (\pi^2+64\kappa^2)^2} \,.
	\label{anomalous_dimension_critical}
\end{equation}
As a consistency check, note that for $\kappa=0$ we recover the results of ref.~\cite{Halperin:1973jh}, while for $\kappa\rightarrow\infty$ we recover the results of refs.~\cite{Ma:1972zz,Ma:1973cmn} for the critical $O(2N_f)$ model.

\section{Fermionic theories}
\label{sec:fermions}

\subsection*{Gross-Neveu QED}
The third theory we consider is Gross-Neveu QED$_3$, the theory of $N_f$ massless Dirac fermions $\psi^m$ ($m=1,\dots,N_f$), coupled to $U(1)_k$ and with the quartic coupling $g$ at the UV fixed point, i.e.~$g=\infty$.
The matter Lagrangian is
\begin{equation}
	\Lgr_{\rm matter}= \bar{\psi}^m\slashed{\covD} \psi^m + \frac{\sigma}{\sqrt{N_f}} \bar{\psi}^m \psi^m + \frac{y}{\sqrt{N_f}}\sigma^3 \,,
\label{grossneveu lagrangian}
\end{equation}
where $\covD_\mu = \partial_\mu + i a_\mu$ is the covariant derivative and $\sigma$ is the HS field with scaling dimension $1+\mathcal{O}(1/N_f)$, whose equation of motion imposes $\bar{\psi}^m \psi^m=0$. We have included in the Lagrangian the classically marginal coupling $y$ (held fixed in the large $N_f$ limit), which is generated for any finite $\kappa\neq 0$. Similarly to the bosonic case, the continuous global symmetry is $SU(N_f) \times U(1)_{\rm m}$, and theory also enjoys parity symmetry if $\kappa=y=0$ and $N_f$ is even. In the limit $\kappa\rightarrow\infty$ we recover the $O(2N_f)$ Gross-Neveu model restricted to the $U(1)$-singlet sector. 

We would like to study the vacuum stability of Gross-Neveu QED$_3$, and to this end we shall compute the effective potential $\Veff$. Similarly to what we did for bosonic QEDs, we perform the computation at the leading order in the $1/N_f$ expansion and to all orders in $\kappa$ and $y$. Therefore, we write $\sigma = \sqrt{N_f} \Sigma +\delta\sigma$, being $\Sigma$ a v.e.v.~that scales as $\mathcal{O}(N_f^0)$, and we path-integrate over $\psi^m$ and $\delta\sigma$. Importantly, in contrast to the bosonic case, for fermionic vector models the HS field has to be path-integrated along the real axis, in order to get a convergent path integral. All in all we get
\begin{equation}
\Veff (\Sigma) = N_f \left( y \Sigma^3 + \frac{1}{6\pi}|\Sigma|^3 \right) \,.
\label{effective_potential_GN}
\end{equation}
In eq.~\eqref{effective_potential_GN} the first term is the classical contribution, whereas the second term comes from quantum fluctuations.  
Due to the real path-integration contour of $\sigma$, the stability of the vacuum at $\Sigma=0$ now requires that
\begin{equation}
\det H = 6\left(y \text{ sign}(\Sigma) + \frac{1}{6\pi}\right) |\Sigma| > 0 \,,
\end{equation}
and therefore we get the stability bound\footnote{This result is consistent with ref.~\cite{Aharony:2018pjn}, where the $\kappa\rightarrow\infty$ limit of the stability bound can be recovered as a particular case of their eq.~(1.10), namely in the limit $|\lambda_F|=1-|\lambda_B|\rightarrow 0$. We find perfect agreement using the dictionary $x_6/\lambda_F=-16\pi y$.}
\begin{equation}
	|y| < \frac{1}{6\pi} \,.
	\label{fermion_bound}
\end{equation}
Again, this condition implies that the local minimum is also a global one, since the second derivative never changes its sign.
As is turns out, a cubic scalar potential for $\Sigma$ -- which would be classically unbounded from below since $\Sigma$ can have both signs, corresponding to the sign of the effective fermion mass -- is allowed by quantum corrections as long as $y$ lies within the region of eq.~\eqref{fermion_bound}.
This is due to the fact that fermionic self-interactions are attractive and tend to stabilize the vacuum, as opposite to what happens in the bosonic case.
In the appendix, we discuss the effective potential and the phases of the theory in presence of a massive and a quartic deformation, showing how the same bound on $y$ holds.

Next, we shall check whether the condition shown in eq.~\eqref{fermion_bound} is satisfied at the RG fixed points of Gross-Neveu QED$_3$. To this end we shall compute the $\beta$ function for $y$. At the leading order in the $1/N_f$ expansion and to all orders in $\kappa$ and $y$ we find \cite{DiPietro:2023gzi}
\begin{align}
\begin{split}
	&\beta_y(y,\kappa) = \\
	& \frac{32}{3\pi^2 N_f} \left( -864 y^3  + \frac{3(4096\kappa^4+640\pi^2\kappa^2-3\pi^4)}{(\pi^2+64\kappa^2)^2} y \right.\\
	& + \left. \frac{4\pi^3(320\kappa^2-3\pi^2)\kappa}{(\pi^2+64\kappa^2)^3} \right)\,.
	\label{betay}
\end{split}
\end{align}
This $\beta$ function is a cubic polynomial in $y$ with no quadratic term, and gauge interactions only affect the linear and constant terms. Since the zeros of $\beta_y$ are odd functions of $\kappa$, we can assume $\kappa>0$ without loss of generality. Before discussing the zeros of $\beta_y$ for generic $\kappa$, let us first study the special cases for $\kappa =0$ and $\kappa\rightarrow\infty$. 

When $\kappa= 0$ from eq.~\eqref{betay} we get
\begin{equation}
	\beta_y(y,0) = \frac{96}{\pi^2 N_f} \left(
	-96 \, y^3 - 3 y\right) \,,
\end{equation}
and so we find only one fixed point corresponding to the real zero at $y^*=0$, which lies in the middle of the stability region of eq.~\eqref{fermion_bound}. Since $\partial_y \beta_y(y^*=0,0)<0$, the operator $\sigma^3$ is relevant here, meaning that tuning of $y$ is needed to reach this fixed point.

In the limit $\kappa\rightarrow\infty$ the $\beta$ function agrees with the findings of ref.~\cite{Aharony:2018pjn} for the ungauged Gross-Neveu model\footnote{In particular, our result \eqref{aharony} matches eq.~(3.65) of ref.~\cite{Aharony:2018pjn}, using the dictionary $N_F=N_f$ and $\lambda^F_6=-16\pi y$.}
\begin{equation}
	\beta_y(y,\infty) = \frac{32}{\pi^2 N_f} \left(
	-288 \, y^3 + y\right) \,.
\label{aharony}
\end{equation}
This $\beta$ function has a fixed point at $y^*=0$ with $\sigma^3$ irrelevant, and two (parity-related) zeros at $y^*=\pm \sqrt{2}/24$ with $\sigma^3$ relevant. Only the fixed point at $y^*=0$ lies within the vacuum stability bound, and corresponds to the usual Gross-Neveu CFT.

\begin{figure}
	\centering
	\hspace{-0cm}\includegraphics[width=0.5\textwidth]{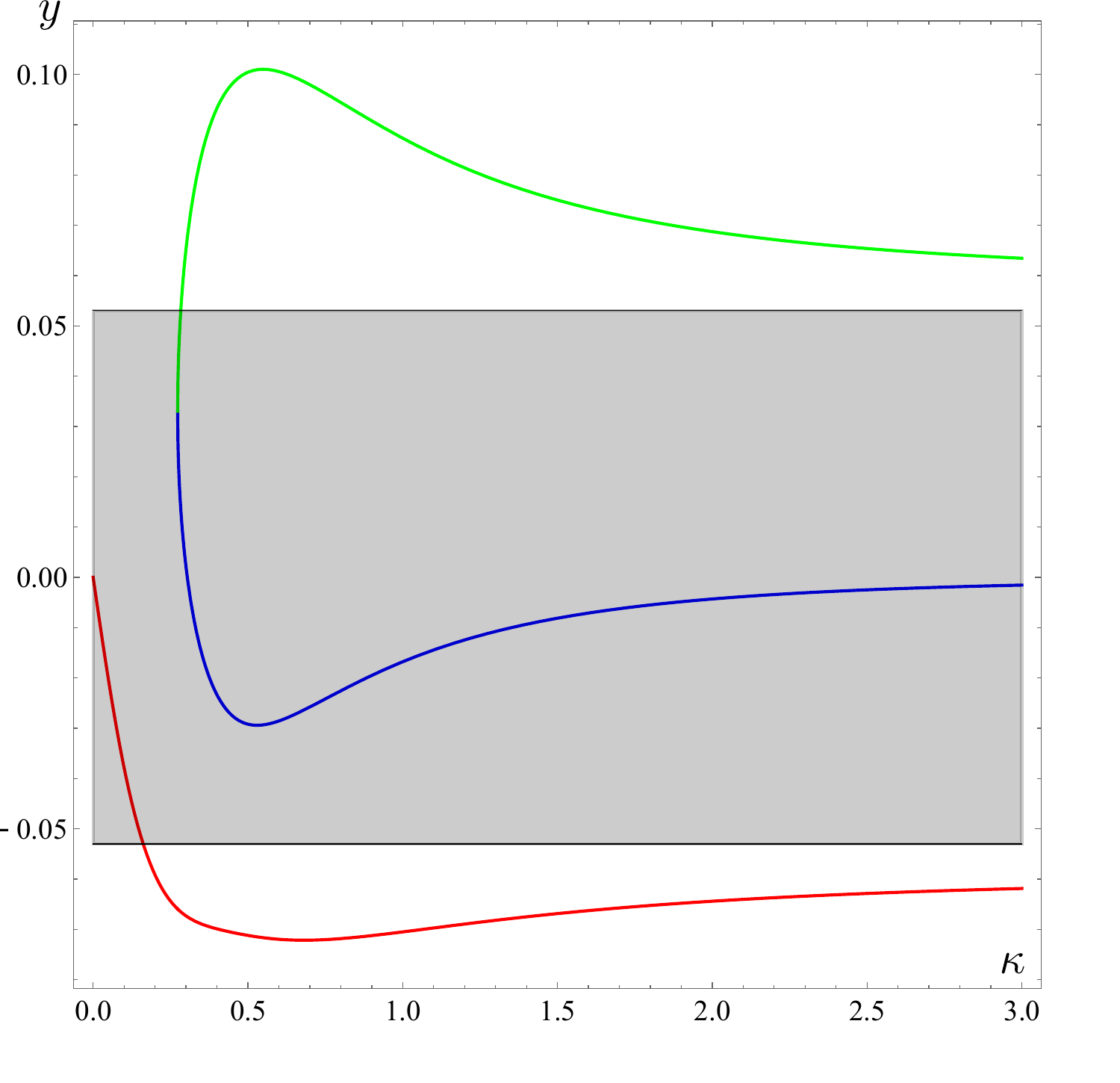}
	\caption{Families of fixed points of $\beta_y$ as a function of $\kappa$ for Gross-Neveu QED$_3$. The fixed points corresponding to stable vacua must lie inside the gray region given by $|y| < \frac{1}{6\pi}$.}
	\label{yfixedpoints}
\end{figure}

In general, we can solve numerically for the zeros of $\beta_y$ as functions of $\kappa$. The results are presented in figure \ref{yfixedpoints}, which shows that there are three families of zeros. A first family of fixed points, depicted by the red curve in the figure, has a relevant $\sigma^3$ operator and exists for any value of $\kappa$, interpolating from $(\kappa,y)=(0,0)$ to $(\kappa,y)=(\infty,-\sqrt{2}/24)$. Two additional families of fixed points are depicted by the blue and green curve in the figure. The operator $\sigma^3$ is irrelevant/relevant there, respectively. They exist for every $\kappa \geq \kappa_1 \simeq 0.273$, below which they annihilate and become complex conjugate, and they reach the values $y=0$ and $y=+\sqrt{2}/24$, respectively, for $\kappa\rightarrow\infty$. The fixed points on the red curve instead are stable only for $\kappa \lesssim 0.162$. The fixed points on the blue curve always lie within the region of vacuum stability. Finally, the fixed points on the green curve have a stable vacuum only if $\kappa_1 \lesssim \kappa \lesssim 0.283$. For $0.162 \lesssim \kappa \lesssim \kappa_1$ there is a `blind spot' where there is no CFT with a stable vacuum.

The anomalous dimension of the lowest-lying singlet $\sigma$ at the leading order in $1/N_f$ and to all orders in $\kappa$ is \cite{DiPietro:2023gzi}
\begin{equation}
	\gamma_{\sigma} = 
	- \frac{16}{3\pi^2 N_f} \frac{9\pi^4-896\pi^2\kappa^2+4096\kappa^4}{ (\pi^2+64\kappa^2)^2} \,,
	\label{anomalous_dimension_criticalGN_ferm}
\end{equation}
which does not depend on $y$. As already observed in ref.~\cite{DiPietro:2020fya} for the special case of $\kappa=0$, this result is identical to that for the anomalous dimension of $\sigma$ in critical bosonic QED$_3$, see eq.~\eqref{anomalous_dimension_critical}, up to $\mathcal{O}(1/N_f^2)$. As a consistency check, in the limit $\kappa\rightarrow \infty$ we indeed recover the anomalous dimension for the HS field of the $O(2N_f)$ Gross-Neveu model \cite{Gracey:1990wi}, while for $\kappa=0$ we recover the results of refs.~\cite{Benvenuti:2019ujm,Boyack:2018zfx}.\footnote{Ref.~\cite{Gracey:1993zn} computed $\gamma_\sigma$ at large $N_f$ as a function of $\kappa$ in the theory with $y=0$. We find a disagreement because in the calculation of the 3-point function $\langle\sigma \bar\psi \psi\rangle$ we include the two-loop diagram with the insertion of $\sigma$ in a closed fermion loop, which is connected to the external fermion line by two photon propagators. This diagram was discarded in \cite{Gracey:1993zn} because it contains the trace of three gamma matrices and the calculation was performed keeping the dimension of spacetime generic. Here we are keeping the dimension fixed to $d=3$ and therefore this trace gives a nonzero contribution. We thank J. Gracey for discussions on this point.}

\subsection*{Fermionic QED}
The fourth theory we consider is fermionic QED$_3$, where the mass of the $N_f$ Dirac fermions is still tuned to zero, but there is no quartic fermionic self-interaction.
The matter Lagrangian is
\begin{equation}
	\Lgr_{\rm matter}= \bar{\psi}^m\slashed{\covD} \psi^m \,,
	\label{fermionicqedlagrangian}
\end{equation}
where $\covD_\mu = \partial_\mu + i a_\mu$ is the covariant derivative. Similarly to the Gross-Neveu case, the continuous global symmetry is $SU(N_f) \times U(1)_{\rm m}$, and theory also enjoys parity symmetry if $\kappa=0$ and $N_f$ is even. In the limit where $\kappa\rightarrow\infty$, it describes $2N_f$ free Majorana fermions restricted to the $U(1)$-singlet sector. Fermionic QED$_3$ has no marginal deformations and it leads to a stable CFT for any $\kappa$, as expected. Indeed, since the large-$N_f$ effective potential does not depend on gauge interactions, the stability analysis is the same as the one for a free fermion theory, which is clearly always stable.

The anomalous dimension of the lowest-lying singlet $\bar\psi^m\psi^m$, at the leading order in $1/N_f$ and to all orders in $\kappa$ reads \cite{DiPietro:2023gzi}
\begin{equation}
	\gamma_{\bar\psi\psi} = 
	 \frac{128}{3N_f}\frac{\pi^2-128\kappa^2}{(\pi^2+64\kappa^2)^2}  \,.
	\label{anomalous_dimension_critical_ferm}
\end{equation}
As a consistency check, in the limit $\kappa\rightarrow \infty$ we get $\gamma_{\bar\psi\psi}=0$, as it should be for a theory of free fermions, while for $\kappa=0$ we recover the result of refs.~\cite{Hermele:2005dkq,Hermele_2007}. Again, we have that at this order of the large-$N_f$ expansion, the fermionic result agrees with the bosonic one, $\gamma_{\bar\psi\psi} = \gamma_{\phi^{\dagger }\phi}$.

\section{Discussion}
\label{sec:discussion}

In this work, we have studied the fixed points of either bosonic or fermionic three-dimensional large-$N_f$ QEDs coupled to a Chern-Simons term at level $k$, with fixed $\kappa=k/N_f$. For each of these theories we computed the effective potential, as well as the $\beta$ functions of the classically marginal couplings, at the leading order in the $1/N_f$ expansion and to all orders in $\kappa$ and in the couplings.
For tricritical bosonic and Gross-Neveu QED$_3$ our analysis shows that a CFT with a stable vacuum can only exist when $\kappa$ lies in the gray region of fig.~\ref{hfixedpoints} and \ref{yfixedpoints}, respectively. There is no such restriction for critical bosonic and fermionic QED$_3$. 

In the paper \cite{DiPietro:2023gzi} we provide details about the computation of the $\beta$ functions, and we couple the 3d large $N_f$ models studied in this work to a 4d real scalar field via a bulk/boundary interaction. This allows us to construct interacting, unitary, and stable conformal boundary conditions for the 4d CFT of a free scalar. 

\section*{Acknowledgements}
\noindent We are grateful to Ofer Aharony and Marco Serone for discussions. We thank J. Gracey for correspondence on the results of \cite{Gracey:1993zn}. LD acknowledges support from the program ``Rita Levi Montalcini'' for young researchers and from the INFN ``Iniziativa Specifica ST\&FI''. EL would like to thank Nikolay Bobev and Balt van Rees for their support. PN would like to thank Thomas Dumitrescu for discussions, and Lorenzo Maffi for a useful conversation on global stability. PN is grateful to ICTP and University of Trieste for the kind hospitality during the preparation of this work. The work of PN is supported by the Mani L.~Bhaumik Institute for Theoretical Physics and by a DOE Early Career Award under DE-SC0020421.

\appendix*
\section{Deformations and phase diagrams}
\label{appendix}
In this appendix, we discuss the phase diagrams of the QED theories described above upon turning on relevant deformations. Since the effective potential does not depend on the gauge interactions at leading order in the large $N_f$ expansion, our discussion is the same as for the case of ungauged vector models, which constitute a particular case of the analysis performed in refs.~\cite{Aharony:2018pjn,Choudhury:2018iwf,Dey:2018ykx}.

\begin{figure}
	\centering
	\hspace{-0cm}\includegraphics[width=0.4\textwidth]{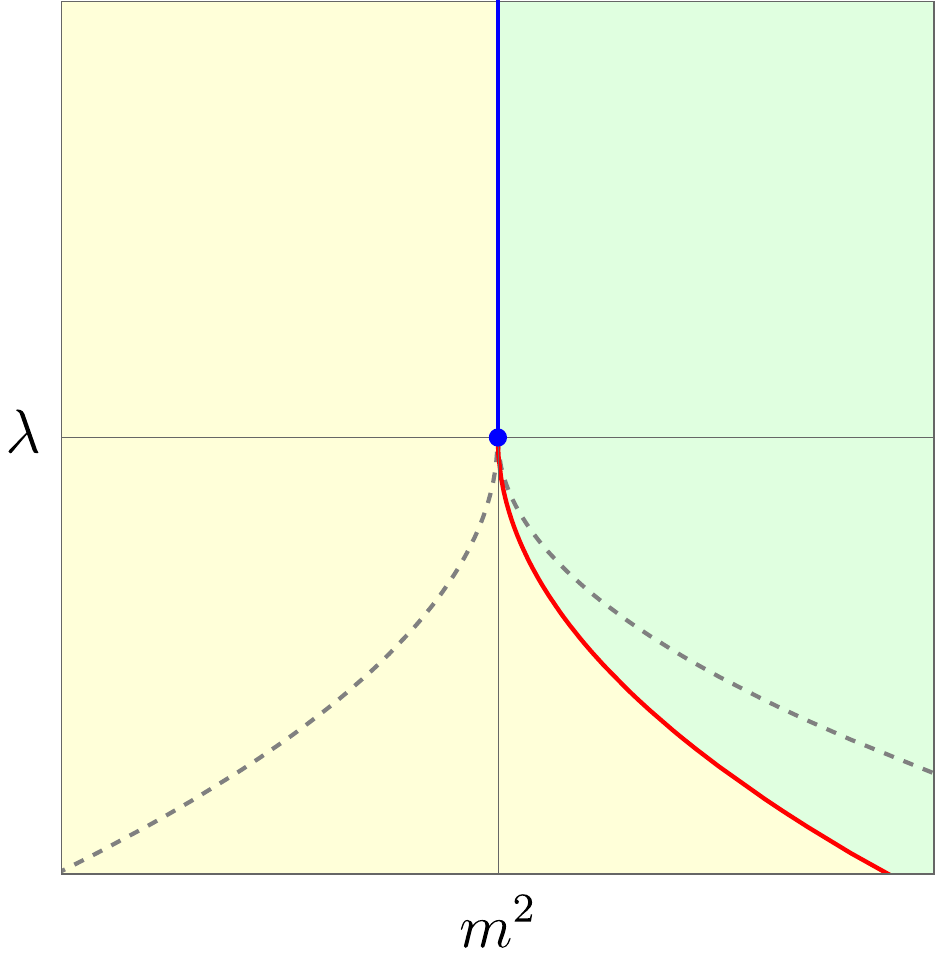}
	\caption{Phase diagram of tricritical QED$_3$ as a function of $\lambda$ and $m^2$, in the range $0<h<16\pi^2/3$. We have depicted in blue the second-order line and the tricritical point. Above the dashed gray lines, there is a unique minimum (an unHiggsed vacuum in green, a Higgsed vacuum in yellow), whereas below such lines the two distinct minima coexist and become degenerate on the first-order line, depicted in red. Its location is specified by $f(h)$, here we have chosen for definiteness a value $h<8\pi^2/3$, so that the transition happens for $m^2>0$.}
	\label{hmiddle}
\end{figure}

Let us first consider the bosonic cases. Tricritical QED$_3$ admits two relevant deformations, the mass $m^2$ and the quartic coupling $\lambda$. The Lagrangian in the presence of such terms reads
\begin{multline}
\Lgr_{\rm matter}= (\covD_\mu \phi^m)^\dagger(\covD_\mu \phi^m) + \sigma \left( \frac{\phi^{\dagger m} \phi^m}{\sqrt{N_f}} - \rho\right) \\ + \frac{h}{\sqrt{N_f}}\,\rho^3 
+ \lambda \, \rho^2 + \sqrt{N_f}m^2\rho \,.
\end{multline}
The effective potential is easily found to be
\begin{equation}
\Veff = 
N_f \left( \Sigma v^2 - \Sigma \eta - \frac{1}{6\pi} \Sigma^{\frac{3}{2}} + h \, \eta^3 + \lambda \, \eta^2 + m^2\, \eta  \right) \,.
\end{equation}
The gap equations for $v^2$ and $\Sigma$
\begin{equation}
2 \Sigma v = 0 \,, \qquad v^2 - \eta - \frac{\sqrt{\Sigma}}{4\pi} = 0\,,
\end{equation}
do not depend on the scalar potential and they can be used to express the effective potential as a function of $\eta$ only:
\begin{equation}
\frac{\Veff}{N_f} = 
\begin{cases}
h \, \eta^3 + \lambda \, \eta^2 + m^2 \, \eta \quad &\text{if } \eta > 0 \,,\\
\left(h-\frac{16\pi^2}{3}\right) \eta^3 + \lambda \, \eta^2 + m^2 \, \eta  \quad &\text{if } \eta < 0 \,.
\end{cases}
\end{equation}
Note that the condition for this potential to be bounded from below gives the same stability bound as in eq.~\eqref{scalar_bound}
\begin{equation}
0 < h < \frac{16 \pi^2}{3} \,.
\end{equation}
As we have previously shown, tricritical QED$_3$ admits an IR fixed point $h^*$ satisfying the stability bound above only if $0 \leq \kappa < \pi/8\sqrt{3}$ and if the flow starts from a UV value of $h$ in the basin of attraction of $h^*$, $h_{\rm green}<h<h_{\rm red}$ (see figure \ref{hfixedpoints}).
From this new $\Veff$, the vacuum is determined by the gap equations for $\eta$, i.e.
\begin{equation}
\begin{cases}
3h \, \eta^2 + 2\lambda \, \eta + m^2 = 0 \quad &\text{if } \eta > 0 \,,\\
(3h-16\pi^2) \eta^2 + 2\lambda \, \eta + m^2 = 0  \quad &\text{if } \eta < 0 \,.
\end{cases}
\label{gapequations}
\end{equation}
From these results we can determine the phase diagram of the theory, which is depicted in figure \ref{hmiddle}. First, let us introduce two useful auxiliary functions of the couplings
\begin{equation}
\begin{split}
\eta_+ &= \frac{-\lambda + \sqrt{\lambda^2-3hm^2}}{3h} \,, \\
\eta_- &= \frac{\lambda - \sqrt{\lambda^2+(16\pi^2-3h)m^2}}{16\pi^2-3h} \,.
\end{split}
\end{equation}

\begin{itemize}
\item If $\lambda \geq 0$ and $m^2 > 0$, there is a unique minimum at $\eta=\eta_-<0$ (unHiggsed).
\item If $\lambda \geq 0$ and $m^2 < 0$ there is a unique minimum at $\eta=\eta_+>0$ (Higgsed).
\item If $\lambda \geq 0$ and $m^2 = 0$ there is a second-order phase transition between the two vacua, which coalesce to $\eta=0$.
\item If $\lambda < 0$ and $m^2 > \frac{\lambda^2}{3h}$, there is a unique minimum at $\eta=\eta_-<0$ (unHiggsed).
\item If $\lambda < 0$ and $m^2 < -\frac{\lambda^2}{16\pi^2-3h}$, there is a unique minimum at $\eta=\eta_+>0$ (Higgsed).
\item If $\lambda < 0$ and $-\frac{\lambda^2}{16\pi^2-3h} \leq m^2 \leq \frac{\lambda^2}{3h}$ (dashed lines in the figure), there are two competing local minima, one at $\eta=\eta_+>0$ and one at $\eta=\eta_-<0$. The two distinct minima become exactly degenerate at $m^2 = \lambda^2 f(h)$, where a first-order phase transition happens.
\end{itemize}
The function $f(h)$ is found to be
\begin{equation}
f(h) = \frac{8\pi^2-3h+8\pi^2\sin\left(\frac{1}{3}\arctan\frac{8\pi^2-3h}{\sqrt{3h(16\pi^2-3h)}}\right)}{3h(16\pi^2-3h)} \,.
\end{equation}
It is straightforward to show that
\begin{equation}
f\left(\frac{16\pi^2}{3}-h\right)=-f(h) \,, \qquad \frac{1}{16\pi^2-3h} \leq f(h) \leq \frac{1}{3h} \,,
\end{equation}
so that $f(h)$ is an odd function with respect to $h=8\pi^2/3$ (where it vanishes), and it is monotonically decreasing from $f(0)=+\infty$ to $f(16\pi^2/3)=-\infty$. Thus, the first-order line ($\lambda<0$, $m^2=\lambda^2 f(h)$) -- depicted in red in the figure -- is at $m^2>0$ ($m^2<0$) if $0<h<8\pi^2/3$ ($8\pi^2/3<h<16\pi^2/3$), and at $m^2=0$ if $h=8\pi^2/3$. It terminates at the tricritical point $(\lambda,m^2)=(0,0)$, where it merges with the second-order line ($\lambda>0$, $m^2=0$) -- depicted in blue in the figure.

Both transitions are between a phase (at $\eta=\eta_-$, green in the figure) described by a $U(1)_k$ Chern-Simons theory with an unbroken $SU(N_f)$ global symmetry, and a phase (at $\eta=\eta_+$, yellow in the figure) where $U(1)_k$ is Higgsed and the $SU(N_f)$ part of the global symmetry is broken down to $SU(N_f-1)$.
The tricritical point is where the tricritical QED$_3$ CFT sits, whereas any point on the second-order line is described by the critical QED$_3$ CFT. Indeed, at low energies, any $\lambda>0$ flows to the critical value $\lambda=\infty$. Note that the location of the phase transitions depends on the regularization scheme (we have used dimensional regularization in this work), but the existence of the various phases does not.

Let us now discuss the phase diagram for critical QED$_3$. The only relevant deformation is the mass term $M \sigma$ (whereas the quartic term $\sigma^2/\lambda$ is obviously irrelevant), and so we consider
\begin{equation}
\Lgr_{\rm matter}= (\covD_\mu \phi^m)^\dagger(\covD_\mu \phi^m) + \frac{\sigma}{\sqrt{N_f}}\,\phi^{\dagger m}\phi^m + \sqrt{N_f }M \, \sigma\,.
\end{equation}
The effective potential reads
\begin{equation}
	\Veff = N_f \left( \Sigma v^2 - \frac{1}{6\pi}\Sigma^{\frac{3}{2}} + M\, \Sigma \right) \,,
\end{equation}
which is always stable, as expected. The gap equations for $v^2$ and $\Sigma$
\begin{equation}
2 \Sigma v = 0 \,, \qquad v^2 - \frac{\sqrt{\Sigma}}{4\pi} + M = 0\,,
\end{equation}
imply that there are two classes of vacua: if $M>0$, then $v^2=0$ and $\Sigma=16\pi^2 M$, so that the theory is in the unHiggsed phase; if $M<0$, then $v^2=-M$ and $\Sigma=0$, so that the theory is in the Higgsed phase. For $M=0$ there is a second-order phase transition where the vacuum is $(v^2,\Sigma)=(0,0)$, and this is described by the critical QED$_3$ CFT. We see that any horizontal line at $\lambda>0$ intersecting the phase diagram of figure \ref{hmiddle} reproduces this one-dimensional phase diagram of critical QED$_3$, as it should be. An alternative way to arrive to the same conclusion is to allow ourselves to use the gap equation for $\Sigma$ to express $\Veff$ as a function of $v^2$ only
\begin{equation}
\frac{\Veff}{N_f} = \frac{16\pi^2}{3}(v^2+M)^3 \,,
\end{equation}
with the condition $v^2 \geq \text{max}(-M,0)$. Note that this potential is always bounded from below, confirming that the vacuum is always stable. 
Interestingly, the upper stability bound that we found for tricritical QED$_3$ amounts to require that the sextic coupling $h$ does not exceed the value it has, effectively, in critical QED$_3$.

\begin{figure}
	\centering
	\hspace{-0cm}\includegraphics[width=0.4\textwidth]{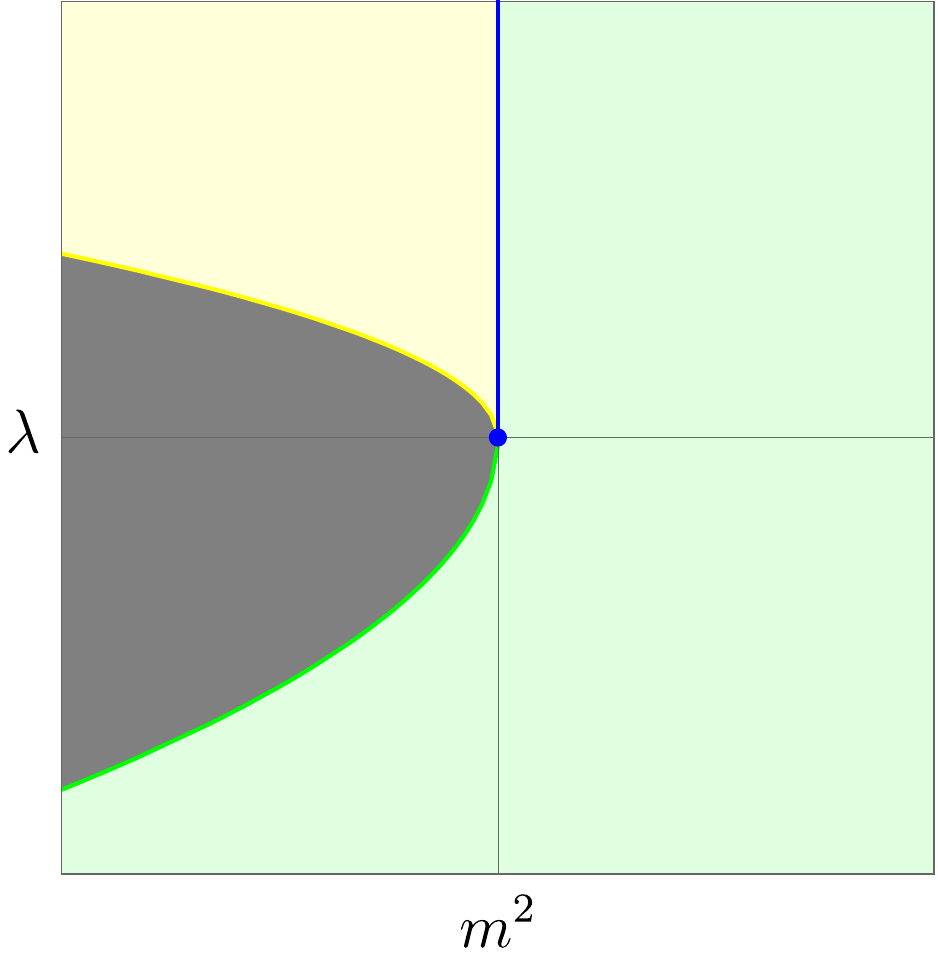}
	\hspace{-0cm}\includegraphics[width=0.4\textwidth]{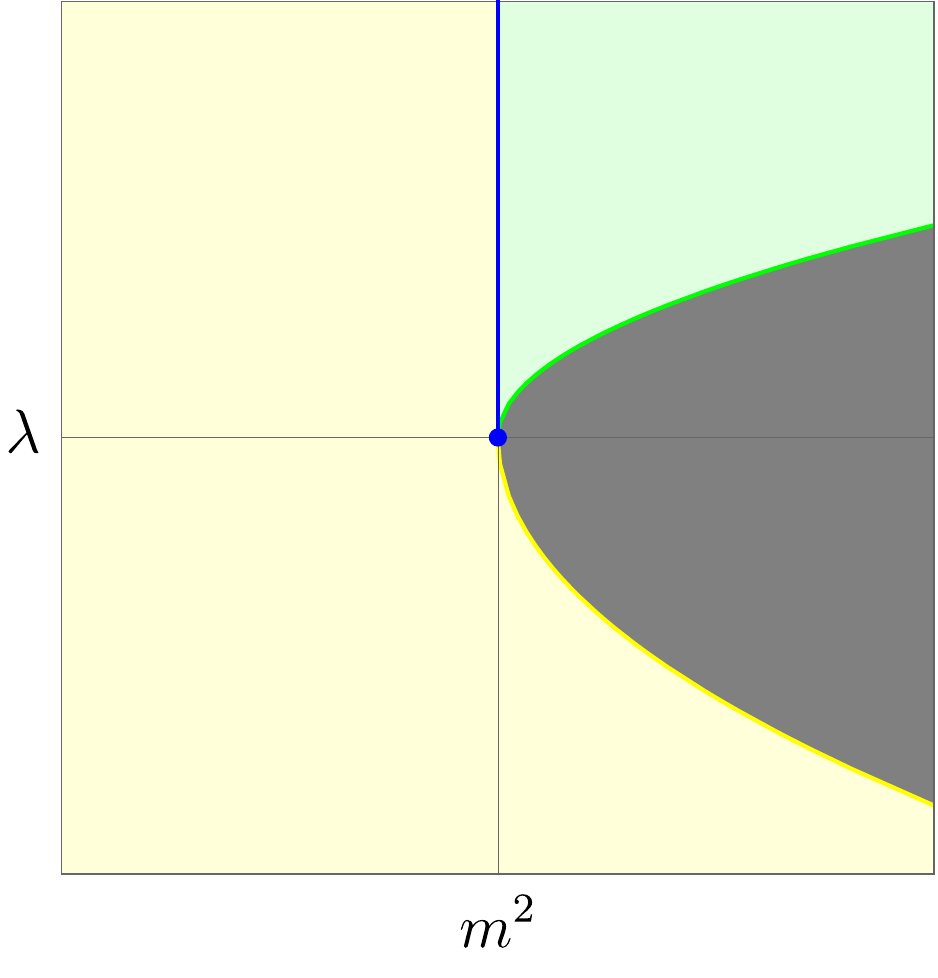}
	\caption{Phase diagram of tricritical QED$_3$ as a function of $\lambda$ and $m^2$, in the range $h<0$ (top) and $h>16\pi^2/3$ (bottom). We have depicted in blue the second-order line and the multicritical point, in green the unHiggsed vacuum, and in yellow the Higgsed vacuum. In the dark gray regions, no solutions to the gap equations exist.}
	\label{hout}
\end{figure}

Finally, let us discuss the phase diagram when we are outside the stability bound. This scenario is depicted in figure \ref{hout}. By investigating the gap equations, one can show that if $h$ is outside the stability interval, then there is a region of the phase diagram where the gap equations \eqref{gapequations} do not have any solution. This region stays in the half-plane of $m^2<0$ ($m^2>0$) for $h<0$ ($h>16\pi^2/3$) and it is delimited by the curves $m^2=\frac{\lambda^2}{3h}$ and $m^2=\frac{\lambda^2}{3h-16\pi^2}$ (yellow and green lines in the figure, respectively).
Moreover, when the gap equations have a solution, this always corresponds to a metastable minimum, since the effective potential is unbounded from below. Curiously, while the first-order transition for $\lambda<0$ disappears, the second-order transition for $\lambda>0$ is still present. We interpret this as the fact that the sextic term in critical QED$_3$ is always (strongly) irrelevant. Hence, starting from tricritical QED$_3$ with any sextic coupling $h$, as soon as we switch on the relevant quartic deformation with positive $\lambda$, at low energies the theory always flows to the critical theory, irrespectively of the initial value of $h$. The metastable vacua of figure \ref{hout} become stable in the limit $\lambda\rightarrow +\infty$, when the runaway behavior is suppressed.

Let us now consider the fermionic case. Gross-Neveu QED$_3$ admits two relevant deformations, the mass $\hat{m}^2$ and the quartic coupling $\hat\lambda$. The Lagrangian reads
\begin{multline}
\Lgr_{\rm matter}= \bar{\psi}^m\slashed{\covD} \psi^m + \frac{\sigma}{\sqrt{N_f}} \bar{\psi}^m \psi^m \\
+ \frac{y}{\sqrt{N_f}}\sigma^3 + \hat\lambda \, \sigma^2 + \sqrt{N_f} {\hat{m}}^2 \, \sigma \,,
\end{multline}
and leads to the following effective potential
\begin{equation}
\frac{\Veff}{N_f}=
\begin{cases}
\left( y + \frac{1}{6\pi} \right) \Sigma^3 + \hat\lambda \, \Sigma^2 + {\hat{m}}^2 \, \Sigma \qquad &\text{if } \Sigma > 0 \,,\\
\left( y - \frac{1}{6\pi} \right) \Sigma^3 + \hat\lambda \, \Sigma^2 + {\hat{m}}^2 \, \Sigma \qquad &\text{if } \Sigma < 0 \,.
\end{cases}
\end{equation}
Note that the condition for this potential to be bounded from below gives the same stability bound as in eq.~\eqref{fermion_bound}
\begin{equation}
|y| < \frac{1}{6\pi} \,.
\end{equation}
As we have previously shown, Gross-Neveu QED$_3$ admits an IR fixed point $y^*$ satisfying the stability bound above only if $\kappa \gtrsim 0.273$ and the flow starts from a UV value of $y$ in the basin of attraction of $y^*$, $y_{\rm red}<y<y_{\rm green}$ (see figure \ref{yfixedpoints}).
The discussion of the phase diagram of Gross-Neveu QED$_3$ is now exactly the same as the one of tricritical QED$_3$, using the dictionary
\begin{equation}
\begin{split}
\eta &\leftrightarrow \frac{\Sigma}{2^{4/3}\pi} \,, \qquad h \leftrightarrow 16\pi^3\left( y + \frac{1}{6\pi} \right) \,, \\
\lambda &\leftrightarrow 2^{8/3}\pi^2 \, \hat\lambda \,, \qquad m^2 \leftrightarrow 2^{4/3}\pi \,{\hat{m}}^2 \,.
\end{split}
\end{equation}
Obviously, now the phases have a different low-energy description. Since the sign of $\Sigma$ gives the sign of the effective fermion mass, phases with positive (negative) $\Sigma$ are characterized by a low-energy $U(1)$ Chern-Simons theory at level $k \pm N_f/2$, respectively, both with an unbroken $SU(N_f)$ global symmetry. The multicritical point at the origin is described by the Gross-Neveu QED$_3$ CFT, whereas each point on the second-order line ($\hat\lambda>0$, ${\hat{m}}^2=0$) is described at low energies by the fermionic QED$_3$ CFT. Indeed, any positive $\hat\lambda$ flows to $\hat\lambda=\infty$ or, equivalently, the quartic fermion coupling $g \sim \hat{\lambda}^{-1}$ flows to $g=0$.

Finally, fermionic QED$_3$ has only one relevant deformation, given by the mass term $m\bar\psi^m\psi^m$ to be added to the Lagrangian \eqref{fermionicqedlagrangian}. Clearly, positive and negative $m$ give rise exactly to the same phases we got in Gross-Neveu QED$_3$ for positive and negative $\Sigma$, respectively.

\bibliography{Letterbib.bib}

\end{document}